\documentclass[12pt, a4paper]{amsart}
\usepackage{amsmath,amssymb,amsfonts}
\usepackage{graphicx}
\usepackage[scaled=1.0]{newtxtext,newtxmath}
\usepackage{hyperref}

\usepackage{fancyhdr}
\usepackage[
	hmarginratio={1:1},     
	vmarginratio={1:1},    
	textwidth=450pt,        
	tmargin=1in,
	heightrounded,         
]{geometry}

\usepackage{bbm}
\usepackage{multirow}
\usepackage{bm}

\usepackage{tikz}
\usetikzlibrary{arrows}
\usetikzlibrary{arrows.meta}
\tikzset{
	tipA/.tip={Triangle[angle=45:10pt]}	
}

\usepackage{makecell}
\usepackage[english]{babel}

\usepackage[colorinlistoftodos]{todonotes}

\usepackage{scalerel}
\newlength\bshft
\def\mybold#1{\ThisStyle{\ooalign{$\SavedStyle#1$\cr%
			\kern-\bshft$\SavedStyle#1$\cr%
			\kern\bshft$\SavedStyle#1$}}}

\newcommand\phase[2]{ \vvmathbb{#1}\vvmathbb{#2} }
\newcommand\phaseol[2]{ \vvmathbb{#1}  \overline{\vvmathbb{#2}} }
\newcommand\phasel[1]{ \vvmathbb{#1}{\ell}}

\title[The phase diagram for mpASEP]
{The phase diagram for a class of multispecies permissive asymmetric exclusion processes}

\author{Dipankar Roy}
\address{Dipankar Roy, Department of Mathematics, Indian Institute of Science\\
	Bangalore - 560012, India.}
\email{dipankarroy@iisc.ac.in}
\date{\today}

\begin{document}

\begin{abstract}
In this article, we investigate a multispecies generalization of the single-species asymmetric simple exclusion process defined on an open one-dimensional lattice. We devise an exact projection scheme to find the phase diagram in terms of densities and currents of all species. In most of the phases, one or more species are absent in the system due to dynamical expulsion. We observe shocks as well in some regions of the phase diagram. We explain the density profiles using a generalized shock structure that is substantiated by numerical simulations.
\end{abstract}

\maketitle

\section{Introduction} 
The asymmetric simple exclusion process (ASEP) is a prominent model for driven diffusion of interacting particles on a lattice. ASEP is often regarded as paradigmatic in the study of nonequilibrium phenomena in statistical physics because of its rich phenomenology \cite{BE2007}. The open single-species ASEP involves particles moving with asymmetric dynamical rules and hard core nearest neighbour interaction along a finite one-dimensional lattice in touch with reservoirs. The totally asymmetric version (TASEP) was exactly solved by Derrida, Evans, Hakim and Pasquier in a seminal work \cite{dehp1993}. They pioneered a novel technique, namely the matrix ansatz, to derive the phase diagram in the nonequilibrium steady state (NESS) by determining macroscopic observables like current and density. Subsequently, these results were extended
to the partially asymmetric cases through computation of the phase diagrams by means of matrix ansatz and corresponding polynomial representation \cite{sasamoto1999, usw2004}.

Besides the single-species cases, two-species ASEPs manifest interesting physical properties. For example, spontaneous symmetry breaking, phase transitions as well as condensation were observed in a class of two-species models \cite{evans-foster1995, arndt-heinzel-rittenberg-1998, arndt-heinzel-rittenberg-1999}. Another important class of two-species ASEPs are the models with semi-permeable boundaries that forbid one of the two species of particles to enter or exit at the boundaries. Phase transitions and other stationary properties were unearthed for totally and partially asymmetric versions of such semi-permeable models \cite{arita2006a, arita2006b, uchiyama2008, ayyer2009, cantini2017}. Moreover, open two-species totally asymmetric models with general bulk and boundary rates were studied in \cite{ayyer2012}. Integrable boundary conditions of two-species ASEPs were completely classified in \cite{crampe-mallick-ragoucy-vanicat-2015}. One such model with integrable boundaries is the \emph{left-permeable} ASEP or \emph{LPASEP} which involves a combination of permeable and semi-permeable boundaries. Matrix ansatz along with the integrability property were used to derive phase diagram and subphases in LPASEP  \cite{ayyer-finn2018a}. Finally, phase diagrams were also investigated for a couple of totally asymmetric two-species models along with an explicit representation for matrix ansatz in \cite{CEMRV2016}.

The generalization of ASEP by considering multiple species have implications both for applied and theoretical investigations. In particular, multispecies ASEPs are useful in the modelling of cell motility \cite{penington2011}, traffic flow \cite{schadschneider-2000, schadschneider2002, karimipour-1999} and biological systems \cite{aghababai-menon-michael-1999, binder-landman-2009, chou-lohse-1999, chowdhury2000, klump-lipowsky-2003}. However, in contrast to the two-species models, there has been rather limited success in finding the phase diagram for open multispecies ASEPs with arbitrary number of species. Almost two decades ago, the structure of the phase diagram was gleaned for such a multispecies model where rates could be chosen from a continuous or discrete distribution \cite{khorrami-karimipour-2000}. Matrix ansatz technique was employed to analyze a multispecies generalization in the context of traffic flow in \cite{karimipour-1999}, as well as multispecies models with ordered sequential and sub-lattice parallel update schemes in \cite{fouladvand-jafarpour-1999}. More recently, the semi-permeable two-species models were generalized by defining a multispecies ASEP model known as \emph{mASEP} in \cite{CGDW2016}. There,  important results related to its NESS were rigorously proved. Later, the phase diagram for mASEP was understood in \cite{ayyer2017} via the \emph{colouring} technique. Integrable boundary conditions were discussed for a class of multispecies models with arbitrary number of species in \cite{CFRV2016}. Densities and currents were determined through matrix product approach for another class of multispecies model in \cite{FRV2018}. In addition, the two-species LPASEP was generalized to a model called \emph{mLPASEP} in \cite{ayyer-finn2018b}. For mLPASEP too, the colouring technique was utilized to compute the phase diagram.

In the present contribution, we study the phase diagram of a multispecies partially asymmetric exclusion process with arbitrary number of species and permeable boundaries. We shall call our model multispecies \emph{permissive} ASEP (or \emph{mpASEP} in short) because both the boundaries permit a species to be replaced with any other species. We construct a colouring scheme that projects the mpASEP onto the open single-species ASEP model. We identify all phases as well as the coexistence regions in the multi-dimensional phase diagram. In the majority of the phases, we observe interesting physical phenomena such as \emph{dynamical expulsion} when particles of one or more species are expelled from the lattice. Nevertheless, corresponding to each species, we find a phase where high density of that particular species can be adopted in the system. We provide explanation to the coarse features of density profiles by means of the generalized shock picture. We note here that we use only colouring argument to compute the macroscopic quantities and the phase diagram without any matrix ansatz. 

The present article is organized in the following way. First, we define the mpASEP in Section \ref{sec:def}. Then we recall the phase diagram and shock picture in the single species ASEP in Section \ref{sec:asep} because our analysis is built on the results for the single-species case. In Section \ref{sec:mpasep}, we present the phase diagram and all densities and currents in each phase. The penultimate section, i.e. Section \ref{sec:shock}, contains discussion regarding the generalized shock picture.

\section{Model definition}
\label{sec:def}
The open ASEP is a continuous time Markov chain defined on a finite lattice of size $L$. At the boundaries, the lattice is attached to particle reservoirs so that particles might enter or leave from the lattice. Each site in the lattice can either be occupied by exactly one particle or be vacant. The particles are biased to move preferentially towards the right. Below, we define a multispecies model which has completely permeable boundaries. Indeed, the boundary interactions make this model distinct from other multispecies models such as mASEP \cite{CGDW2016, ayyer2017} and mLPASEP \cite{ayyer-finn2018b} where passage of particles is far more restricted at the boundaries.

The precise definition of multispecies permissive ASEP or mpASEP is as follows. Similar to the open ASEP, the mpASEP is defined on an open one-dimensional lattice of size $L$. There are total of  $(r+1)$ species of particles in the mpASEP. Each species is identified with an element of the label set $\vvmathbb{L} := \left\lbrace 0,1,\ldots, r\right\rbrace $. The preference of rightwards hopping is determined by the natural order relation among the species: $0<1< \cdots< r$. One can think of 0's as vacancies or the slowest species, $r$'s as the fastest species, and any intermediate species $i \ (0<i<r)$ being faster or slower than another species $j$ according to where $i>j$ or $i<j$ respectively. The left boundary interactions are given by
\begin{equation}
	i \rightarrow j  \text{ with rate } 
	\begin{cases}
		\alpha_{j} \text{ if } i<j, \\
		\gamma_{j} \text{ if } i>j,
	\end{cases}
 \end{equation}  
where $(\alpha_{1}, \ldots , \alpha_{r}) $ and $(\gamma_{0}, \ldots , \gamma_{r-1}) $ are sets of fixed positive parameters. In the bulk exchange of particles between neighbouring sites follows the rule
\begin{equation}
	ij \rightarrow ji  
	\text{ with rate } \begin{cases}
		1 \text{ if } i>j, \\
		q \text{ if } i<j,
	\end{cases}
\end{equation}  
where we impose the condition $0< q<1$. On the right boundary particles can be replaced with the following rates
\begin{equation}
	i \rightarrow j  \text{ with rate } 
	\begin{cases}
		\beta_{j} \text{ if } i>j, \\
		\delta_{j} \text{ if } i<j, 
		
	\end{cases}
\end{equation}  
where the parameters $(\beta_{0}, \ldots , \beta_{r-1})$ and $(\delta_{1}, \ldots , \delta_{r})$ are again fixed and positive. Finally, it is useful to introduce $A_{k}:=  \sum_{i=k}^{r} \alpha_{i}$, and $G_{k} := \sum_{i=0}^{k-1} \gamma_{i}$. Also, we define $B_{k} := \sum_{i=0}^{k-1} \beta_{i}$, and $D_{k} := \sum_{i=k}^{r} \delta_{i} $ in terms of the right boundary parameters. These quantities will appear as boundary parameters when we apply the colouring technique.

\subsubsection*{Remark~1} We can define a totally asymmetric variant of the mpASEP with $r$ species in the following manner. We allow $q=0$ as well as $\gamma_{i}=\delta_{j}=0$ for $0 \leq i<r$ and $1 \leq j \leq r$. Moreover, we put restrictions on the nonzero boundary rates such that $\sum_{i=1}^{r} \alpha_{i} \leq 1 $ and $\sum_{i=0}^{r-1} \beta_{i} \leq 1 $. As an example, one checks that the model becomes the single-species TASEP for $r=1$.

 \begin{figure}[htbp!]
 	\begin{center}	
 		\begin{tikzpicture}[scale=0.5]
 		\draw[-{Latex[length=3.5mm]},thick] (0,0)--(0,10);
 		\draw[-{Latex[length=3.5mm]},thick] (0,0)--(10,0);
 		\draw[black, thick] (0,0) rectangle (4.5,4.5);
 		\draw[black, line width = 1.1pt] (4.494,4.494)--(10,10);
 		
 		\node at (2.25, 2.25){MC};
 		\node at (2.25, 7.50){HD};
 		\node at (7.50, 2.25){LD};
 		
 		\node at (10.5,0){$ a $}; 
 		\node at (0,10.5){$ b $}; 
 		\node at (11.2, 10.5){$ a = b $};
 		
 		\node at (0,-0.55){$0$};
 		\node at (-0.55,0){$0$};
 		\node at (4.5,-0.55){$1$};
 		\node at (-0.55,4.5){$1$}; 
 		\end{tikzpicture}
 	\end{center}
 	\caption{The phase diagram for ASEP. The HD-LD coexistence region is the semi-infinite line $1<a=b$.}
 	\label{fig:pdasep}
 \end{figure}
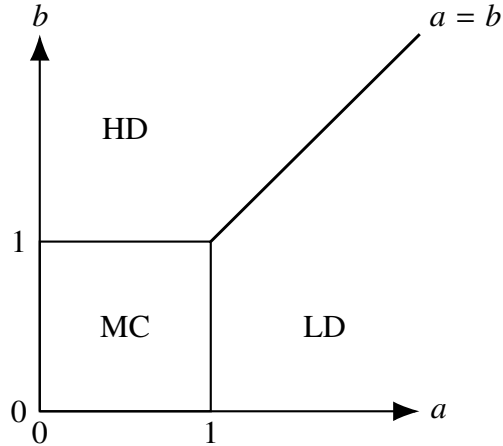

\section{The single-species ASEP}\label{sec:asep}
The open single-species ASEP (or simply ASEP for brevity) is a well-known model \cite{dehp1993,sasamoto1999, usw2004,BE2007}. Here, we review only the important facts that would be relevant to our discussion later. If we let $r=1$ in the mpASEP, then the model is nothing but the ASEP. In the ASEP, there are particles of only one species, denoted with 1, and vacancies, represented by 0. At the left boundary, the injection and removal of particles take place with the following rates
\begin{equation}
	0 \rightarrow 1  \text{ with rate } \alpha, \qquad 1 \rightarrow 0  \text{ with rate } \gamma. \nonumber
\end{equation}  
In the bulk, the rates for hopping forward and backward are
\begin{equation}
	10 \rightarrow 01  \text{ with rate } 1, \qquad  01 \rightarrow 10  \text{ with rate } q. \nonumber
\end{equation}  
Additionally, the right boundary interactions are given by the rules
\begin{equation}
	1 \rightarrow 0  \text{ with rate } \delta, \qquad  0 \rightarrow 1  \text{ with rate } \beta. \nonumber
\end{equation}  
The parameters $ \alpha, \gamma ,  \beta $ and $\delta $ are positive real numbers.

\begin{table}[htbp!]
	\renewcommand*{\arraystretch}{1.6}
	\begin{tabular}{|c|c|c|c|}
		\hline	
		Phase & Phase Region &  $\rho_{1}$&   $J_{1}$ \\
		\hline
		HD (High Density) &	$\max \left\lbrace  a ,1 \right\rbrace < b $ & $\frac{b}{1+b}$ &   $ (1-q) \frac{b}{(1+b)^2} $\\ \hline
		LD (Low Density) &  $\max \left\lbrace b,1 \right\rbrace < a $ & $\frac{1}{1 + a}$ & $ (1-q) \frac{ a}{ (1+  a)^2} $ \\ \hline
		MC (Maximal Current) &  $\max \left\lbrace  a ,b \right\rbrace < 1 $  &  $\frac{1}{2}$ &  $\frac{1}{4}(1-q)$ \\ 
		\hline 
		HD-LD coexistence line & $ 1< a=b $ & $\frac{1-x+x b}{1+b}$ &    $ (1-q) \frac{b}{ (1+ b)^2} $\\
		\hline
	\end{tabular} 
	\vspace{0.4cm}
	\caption{Currents and bulk densities for the ASEP. We note here that $\rho_{0} + \rho_{1} = 1$, and $J_{0} = -J_{1}$.}
	\label{table:cdasep}
\end{table}

In the thermodynamic limit $L\rightarrow \infty$ \cite{usw2004}, the model exhibits three phases - \emph{high density} or HD, \emph{low density} or LD and \emph{maximal current} or MC, as well as HD-LD \emph{coexistence line} where one observes a shock in the system. The phase regions and bulk density and current in each phase are listed in the Table \ref{table:cdasep} in terms of the boundary parameters $a= \kappa^{+}_{\alpha,\gamma}$ and $b=\kappa^{+}_{\beta,\delta}$, where 
\begin{equation}
	\kappa^{\pm}_{u,v} := \frac{ 1- q - u + v \pm \sqrt{ (1- q -u + v)^{2} + 4 u v}}{2 u}.
\end{equation}
To summarize, the 1's have high density ($>1/2$) in HD phase, low density ($<1/2$) in LD phase, and density of $1/2$ in MC phase. On the coexistence line, the density has a linear profile. $J_{1}$ is $(1-q)/4$ in MC phase, and depends on $a$ (resp. $b$) in LD (resp. HD) phase as given in Table~\ref{table:cdasep}. Figure~\ref{fig:pdasep} shows the structure of the phase diagram which is drawn using $a$ and $b$.

\begin{figure}
	\begin{center}	
		\begin{tikzpicture}[scale=0.5]
		\draw[black,thick] (0,0) rectangle (10,10);
		\draw[black,thick] (0,3)--(6,3)--(6,7)--(10,7);
		
		\node[] at (-1.5,3){$\scriptstyle 1/(1+b)$};
		\node[] at (11.5,7){$\scriptstyle b/(1+b)$}; 
		
		\node[] at (7.5,5){$1$};\node[] at (3,5){$0$};
		\draw[<->] (5.5,5)--(6.5,5);
		
		\node at (-1.2,5){$\rho$};
		\node[] at (0,-0.5){$0$};\node[] at (10,-0.5){$1$};
		\node[] at (-0.5,0){$0$}; \node[] at (10.55,10){1};
		\node[] at (5,-1){$x$};
		\end{tikzpicture}
	\end{center}
	\caption{The shock picture for the ASEP on the coexistence line $1< a = b $. The normalized site position $x$ equals $i/L$. }
	\label{fig:asepshock}
\end{figure}
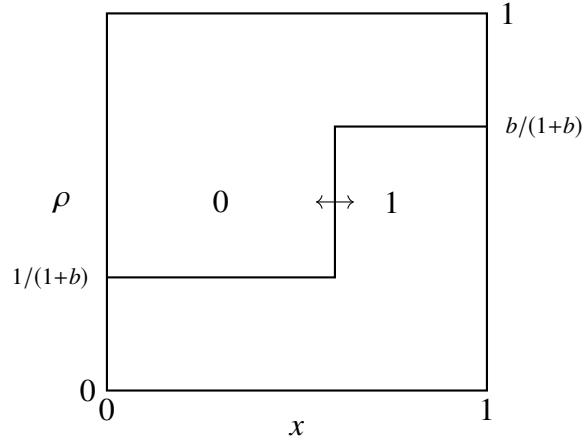

The macroscopic features of the density profiles in different phases can be understood by appealing to a shock picture on the coexistence line $1<a=b$ in the phase diagram. On the coexistence line, the instantaneous density profiles show a sharp discontinuity which is known as a shock. The shock picture describes the shock by plotting first density of the 1's at each normalized position of the lattice, and then density of the 0's on top of it. The shock picture for the ASEP is drawn in Figure \ref{fig:asepshock}. The shock picture illustrates the shock between 1's and 0's by showing two regions, one marked with 1 and another with 0. In every shock picture such as this, height of a region at a site position $x$ equals density of the species that labels the region.

On the coexistence line, the shock formed between 1's and 0's perform a random walk with zero net drift in the system leading to a linear density profile. In the HD phase, the shock acquires negative drift. Thus the shock gets pinned to the left and high density of 1's prevail on the system. In the LD phase, the shock is pinned to the right because it has positive drift. As a result, the system assumes low density. However, as one approaches the MC phase along the coexistence line, the height of the shock becomes zero with both particles and vacancies attaining equal bulk density. In each case, it can be checked that the bulk densities in Table \ref{table:cdasep} are consistent with the explanation provided here in terms of the shock picture in Figure \ref{fig:asepshock}.

\section{The exact phase diagram for the mpASEP}\label{sec:mpasep}
We derive the phase diagram for the mpASEP by constructing projections onto the single-species ASEP. We will refer to this projection procedure as \emph{k-colouring}. The colouring technique was used earlier to study two-species TASEP models in \cite{ayyer2009, ayyer2012, CEMRV2016} as well as to obtain the phase diagrams for mASEP \cite{ayyer2017} and mLPASEP \cite{ayyer-finn2018b}.

The essence of the colouring argument is that a set of particles might appear indistinguishable from the perspective of another set of particles because of the dynamical rules of the process. Thus an appropriate colouring, i.e. identifying particles of different species as a single species, can project the original model onto another model with fewer number of species. We refer to the colouring as exact if it is valid both in the bulk and the boundaries. For such a colouring procedure applicable to the mpASEP, we define $k$-colouring as follows. We label every species $i<k$ with $0_{k}$, and all other species $i \geq k$ with $1_{k}$. Consequently, mpASEP now resembles the ASEP where particles with label $1_{k}$ (resp. $0_{k}$) play the role of particles (resp. vacancies). Before we go into more details for arbitrary $r$, let us first understand the colouring as well as the resulting phase diagram for the simplest nontrivial case, i.e. the case $r=2$, in the next section. We postpone our discussion regarding the general case until Section~\ref{ssec:mpasep}.

\subsection{Phase diagram of mpASEP with 2 species} If we set $r=2$, then mpASEP involves only two types of particles (and vacancies). For the sake of completeness, we note all boundary interactions below:

\begin{center}
	\begin{tabular}{ll}
		\underline{Left}:
		&
		$\begin{cases}
			0, 1 \rightarrow 2 \quad \text{ with rate } \alpha_{2},	\\
			0 \rightarrow 1 \quad \text{ with rate } \alpha_{1},    \\
			1,2 \rightarrow 0 \quad \text{ with rate } \gamma_{0},  \\
			2 \rightarrow 1 \quad \text{ with rate } \gamma_{1}, \\
		\end{cases}$\\
		&\\
		\underline{Right}:
		&$
		\begin{cases}
			1, 2 \rightarrow 0 \quad \text{ with rate } \beta_{0},  \\
			2 \rightarrow 1 \quad \text{ with rate } \beta_{1}, \\
			0, 1 \rightarrow 2 \quad \text{ with rate } \delta_{2},	\\
			0 \rightarrow 1 \quad \text{ with rate } \delta_{1}.    \\
		\end{cases}$
	\end{tabular}
\end{center}
There are two possible colourings. In the first colouring, which we call 1-colouring, we identify 1's and 2's as $1_{1}$, and 0's as $0_{1}$. Then the 2-species mpASEP behaves exactly as the ASEP with the boundary interaction rates given by
\begin{center}
	\begin{tabular}{ll}
		\underline{Left}:
		&
		$\begin{cases}
			0_{1} \rightarrow 1_{1} \quad \text{ with rate } (\alpha_{1}+\alpha_{2}) ,	\\
			1_{1} \rightarrow 0_{1} \quad \text{ with rate } \gamma_{0} ,  \\
		\end{cases}$\\
		&\\
		\underline{Right}:
		&$
		\begin{cases}
			1_{1} \rightarrow 0_{1} \quad \text{ with rate } \beta_{0} ,	\\
			0_{1} \rightarrow 1_{1} \quad \text{ with rate } (\delta_{1} + \delta_{2}) .  \\
		\end{cases}$
	\end{tabular}
\end{center}
The relevant left and right boundary parameters are $a_{1} = \kappa_{\alpha_{1} + \alpha_{2} , \gamma_{0} }^{+}$ and $b_{1} = \kappa_{ \beta_{0} , \delta_{1} + \delta_{2} }^{+}$ respectively. The other possibility is 2-colouring in which we label species 2 with $1_{2}$, and both species 0 and 1 with $0_{2}$. Again the resulting model is the ASEP but with different boundary rates, 
\begin{center}
	\begin{tabular}{ll}
		\underline{Left}:
		&
		$\begin{cases}
		0_{2} \rightarrow 1_{2} \quad \text{ with rate } \alpha_{2} ,	\\
		1_{2} \rightarrow 0_{2} \quad \text{ with rate } (\gamma_{0} + \gamma_{1}) ,  \\
		\end{cases}$\\
		&\\
		\underline{Right}:
		&$
		\begin{cases}
		1_{2} \rightarrow 0_{2} \quad \text{ with rate } (\beta_{0} + \beta_{1}) ,	\\
		0_{2} \rightarrow 1_{2} \quad \text{ with rate }  \delta_{2} .  \\
		\end{cases}$
	\end{tabular}
\end{center}    
The phase diagram for the model just described is determined by the left and right boundary parameters $a_{2} = \kappa_{ \alpha_{2} , \gamma_{0} +\gamma_{1} }^{+}$ and $b_{2} = \kappa_{ \beta_{0} + \beta_{1} , \delta_{2} }^{+}$ respectively. \\

\begin{figure}[htbp!]
	\begin{center}
		\begin{tikzpicture}[scale=0.35]
		\draw[-{Latex[length=3.5mm]},thick] (0,0)--(0,20);
		\draw[-{Latex[length=3.5mm]},thick] (0,0)--(20,0);
		\draw[black,thick] (0,0) rectangle (12.8,7.25);
		\draw[black, thick] (0,0) rectangle (6.9,12.2);
	    \draw[black, line width = 1.1pt] (12.8,7.25)--(20,11.328125);
		\draw[black, line width = 1.1pt] (6.9,12.2)--(11.31147541,20);
		\draw[dashed, black, line width = 0.7pt] (6.9,7.25)--(19.03448276,20);
		
		\node at (22.2, 11.8) {$a_{1} = b_{1}$};
		\node at (21.23, 20.7) {$a_{2} = b_{1}$};
		\node at (11.31147541, 20.7) {$a_{2}=b_{2} $};
		
		\draw[dotted, black] (0, 0)--(6.9, 12.2);
		\draw[dotted, black] (0, 0)--(12.8, 7.25);
		
		\node at (0,-0.75) {0};\node at (-0.75,0) {0};
		\node at (20.75,0) {$a$};\node at (0,20.8) {$b$};
		\node at (7,-0.8) {$L_{2}$}; 
		\node at (12.9,-0.8) {$L_{1}$};
		
		\node at (-0.9, 7.25) {$R_{1}$};
		\node at (-0.9, 12.2) {$R_{2}$};
		
		\node at (3.4, 3.5) {$ \phase{0}{0}$};
		\node at (10, 3.5 ) {$ \phase{0}{1}$};
		\node at (16.5, 3.5 ) {$ \phase{0}{2} $};
		\node at (3.4, 9.75) {$ \phase{1}{0}$};
		\node at (3.4, 16.1) {$ \phase{2}{0}$};
		\node at (12.85, 15.4) {$\phase{1}{1}$};
		\end{tikzpicture}
		
		\caption{The phases of the mpASEP for $r=2$ visualized on a two-dimensional plane that is embedded in the $(a_{1},a_{2},b_{1},b_{2})$-space and determined by parameters $(\mu_1,\nu_{1})$ as shown in \eqref{eq:2species-phase-plane}. The coordinates of the points $L_{i}$ and $R_{i}$ are given by \eqref{eq:lrpoints} in Section~\ref{ssec:mpasep} for the general case.}
		\label{fig:2sp_pd}
	\end{center}
\end{figure}
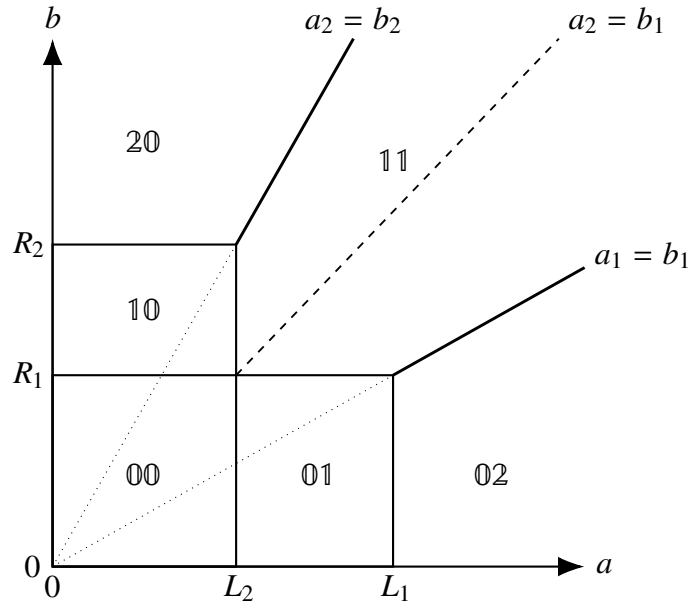

All phases reside in the $(a_{1},a_{2},b_{1},b_{2})$-space. However, it is easy to check that the relevant boundary parameters obey the restrictions $a_{1} < a_{2}$, and $b_{2} < b_{1}$. These constraints allow us to visualize the phases on a two-dimensional plane that is fixed by 
\begin{equation}
	\begin{aligned}
		a_{2}  &=  \mu_{1} a_{1}, \\
		b_{1}  &=  \nu_{1} b_{2}, \\
	\end{aligned} \label{eq:2species-phase-plane}
\end{equation} 
where $\mu_{1}, \nu_{1} > 1$. The parameter $a = \left( a_{1}^{2} + a_{2}^{2}\right)^{1/2}$ gives the radial distance from the origin in the $(a_{1}, a_{2})$-subspace. Similarly, we define $b = \left( b_{1}^{2} + b_{2}^{2}\right)^{1/2}$ in the $(b_{1}, b_{2})$-subspace.

\begin{figure}[htbp!]
	\centering
	\includegraphics[width = 1\textwidth]{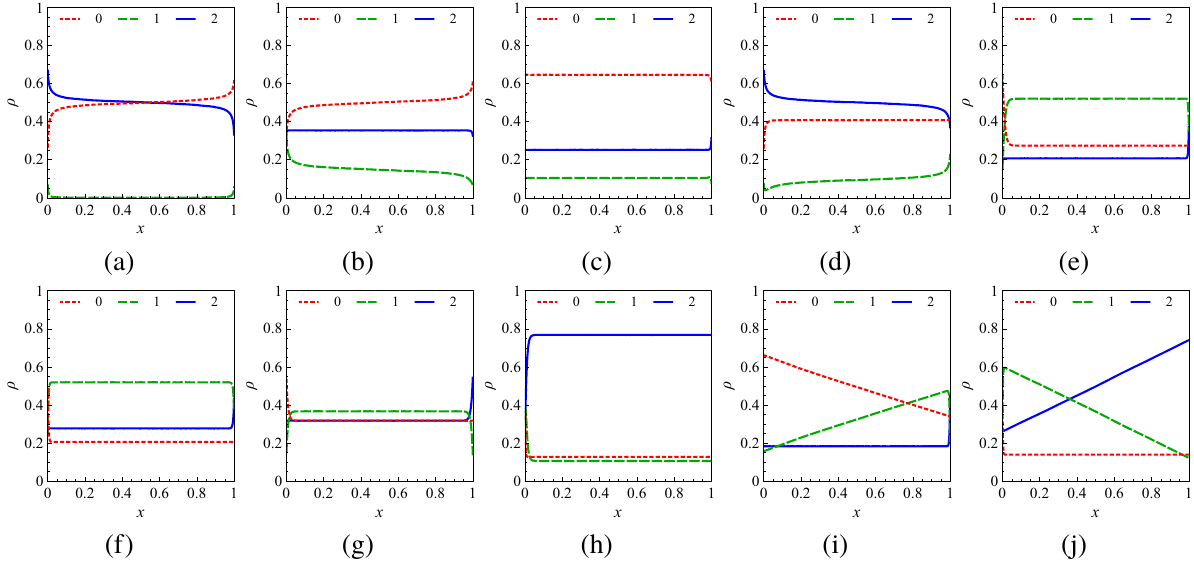}
	\caption{The time-average densities in the two-species mpASEP for species 2 (blue dotted line), 1 (green solid line), and 0 (red dashed line) for (a) phase $\phase{0}{0}$ ($a_{1} \simeq 0.29,a_{2} \simeq 0.45,b_{1} \simeq 0.61 ,b_{2} \simeq 0.47$, (b) phase $\phase{0}{1}$ ($a_{1} \simeq 0.43,a_{2} \simeq 1.83,b_{1} \simeq 0.63 ,b_{2} \simeq 0.48$), (c) phase $\phase{0}{2}$ ($a_{1} \simeq 1.82,a_{2} \simeq 2.98,b_{1} \simeq 0.66,b_{2} \simeq 0.51 $), (d) phase $\phase{1}{0}$ ($a_{1} \simeq 0.29 ,a_{2} \simeq 0.46,b_{1} \simeq 1.46,b_{2} \simeq 0.56$), (e) phase $\phase{1}{1}$ with $b_{1}<a_{2}$ ($a_{1} \simeq 2.04 ,a_{2} \simeq 3.83,b_{1} \simeq 2.66,b_{2} \simeq 0.72 $), (f) phase $\phase{1}{1}$ with $a_{2}<b_{1}$ ($a_{1} \simeq 1.39,a_{2} \simeq2.62 ,b_{1} \simeq 3.87,b_{2} \simeq 0.86$), (g) phase $\phase{1}{1}$ with $a_{2}=b_{1}$ ($a_{1} = 1.27 ,a_{2} = b_{1} =2.15 ,b_{2} =1.33 $), (h) phase $\phase{2}{0}$ ($a_{1} \simeq 1.21 ,a_{2} \simeq 2.64, b_{1} \simeq6.87 , b_{2} \simeq 3.3 $), (i) $\phase{0}{1}-\phase{1}{1}$ coexistence region ($a_{1} \simeq 1.95 ,a_{2} \simeq 4.48 ,b_{1} \simeq 1.95 ,b_{2} \simeq 1.16 $), and (j) $\phase{1}{1}-\phase{2}{0}$ coexistence region ($a_{1} \simeq 1.1, a_{2} \simeq 2.84,b_{1} \simeq 6.21,b_{2} \simeq 2.84 $). The lattice size is 1000 in each case.}
	\label{fig:fig1}
\end{figure}

We derive the phase diagram shown in Figure~\ref{fig:2sp_pd} taking into account all possible colourings. The phase diagram consists of the following phases.
\begin{itemize}
	\item Phase $\phase{0}{0}$: $\max \left\lbrace a_{1}, a_{2}, b_{1}, b_{2} \right\rbrace  < 1$.	
	
	\item Phase $\phase{0}{1}$: $\max\left\lbrace  a_{1}, b_{1} \right\rbrace <1 $, $\max\left\lbrace  1, b_{2} \right\rbrace < a_{2} $.
	
	\item Phase $\phase{0}{2}$: $\max\left\lbrace  1, b_{1} \right\rbrace < a_{1} $.
	
	\item Phase $\phase{1}{0}$: $\max\left\lbrace  a_{1}, 1 \right\rbrace < b_{1} $, $\max\left\lbrace a_{2}, 		 b_{2}\right\rbrace < 1 $.
	
	\item Phase $\phase{1}{1}$: $\max\left\lbrace 1, a_{1}, b_{2} \right\rbrace < \min \left\lbrace  a_{2}, b_{1} \right\rbrace $.
	
	\item Phase $\phase{2}{0}$: $\max \left\lbrace a_{2}, 1\right\rbrace < b_{2} $.
\end{itemize}

Here, we refer to each phase using a two-letter word of the form $\phase{i}{j}$. Our convention is such that the word $\phase{i}{j}$ signifies the following. The phase $\phase{i}{j}$ is mapped to the HD phase of the ASEP by $k$-colouring for $1 \leq k\leq i$. Moreover, $k$-colouring projects onto the LD phase for $ 2-j < k \leq 2 $, and onto the MC phase for $i < k \leq 2-j$. For example, the phase $\phase{1}{0}$ is projected onto the HD phase by $1$-colouring, and onto the MC phase by $2$-colouring. The word $\phase{1}{0}$ indicates the same by our convention.


The currents in these phases are as follows. First, we note that the relation $J_{2} = - J_{1} - J_{0}$ is satisfied for all phases. This is true because the 2's can replace or be replaced by species $0$ and $1$ at the boundaries. Species 2 (resp. 0) has positive (resp. negative) current in all phases. Species 1 has zero current in phase $\phase{0}{0}$ and on the semi-infinite line $1<a_{2} = b_{1}$. $J_{1}$ is positive or negative according to whether $a_2 < b_1$ or $a_{2}>b_{1}$.

The densities can be easily computed as well. Here, we describe only the densities of species 2 and 1 because we have $\rho_{0} + \rho_{1} + \rho_{2} = 1$ in each phase. The 2's have high density in phase $\phase{2}{0}$, and low density in phases $\phase{1}{1}$ and $\phase{0}{2}$. But, $\rho_{2}$ remains constant at 1/2 throughout the phases $\phase{0}{0}$ and $\phase{1}{0}$. Species $1$ is expelled from the lattice in phase $\phase{0}{0}$ even though injection rates ($\alpha_{1}$ and $\delta_{1}$) of the 1's are positive. Such a phenomenon which was observed earlier for multispecies models like mASEP and mLPASEP is referred to as dynamical expulsion \cite{ayyer2017, ayyer-finn2018a}. Moreover, the 1's exhibit low density in all other phases except the phase $\phase{1}{1}$. The 1's can assume either high or low density in phase $\phase{1}{1}$. However, $\rho_1$ varies continuously in this phase. One observes linear density profiles for species 2 and 1 (resp. 1 and 0) on the coexistence line $1< a_2 = b_2 $ (resp. $1< a_1 = b_1$) between the phases $\phase{2}{0}$ and $\phase{1}{1}$ (resp. $\phase{2}{0}$ and $\phase{1}{1}$).

For the exact expression for the densities and currents, the reader is referred to Table~\ref{table:cdmpasep}. The simulation results for densities are shown in Figure~\ref{fig:fig1}.

\subsubsection*{Remark~2} We consider following specializations of the boundary parameters: (1) $ \alpha_{2} = \alpha = 1 - \alpha_{1}, \beta_{0} = \beta = 1 - \beta_{1} , \gamma_{0} = \gamma_{1} = \delta_{2} =\delta_{1} = 0 $,  and (2) $ \alpha_{2} = \alpha = 1 - \alpha_{1}, \beta_{0} = \beta , \gamma_{0} = \gamma_{1} = \delta_{2} =\delta_{1} =  \beta_{1} = 0$, with $\alpha, \beta <1$. Then, if one permits $q=0$, the 2-species mpASEP exactly corresponds to the couple of 2-species TASEP models discussed in \cite{CEMRV2016}. There the models were referred to as $P_{1}$ and $P_{2}$ for specializations (1) and (2) respectively. One can easily check that in the limit $q \rightarrow 0$ the phase diagram in Figure~\ref{fig:2sp_pd} translates to the phase diagrams for $P_{1}$ and $P_{2}$ described in \cite{CEMRV2016} for these specializations.

\subsection{Phase diagram for mpASEP \label{ssec:mpasep}}
The phase diagram for the mpASEP with $r$ species is composed of total $r(r+1)/2$ phases in a $2r$-dimensional space. There are $2r$ independent parameters because of $r$ possible $k$-colourings. On application of each $k$-colouring to the mpASEP, the boundary interactions become 
\begin{center}
	\begin{tabular}{ll}
		\underline{Left}:
		&
		$\begin{cases}
		0_{k} \rightarrow 1_{k} \quad \text{ with rate } A_{k} ,	\\
		1_{k} \rightarrow 0_{k} \quad \text{ with rate } G_{k} ,  \\
		\end{cases}$\\
		&\\
		\underline{Right}:
		&$
		\begin{cases}
		1_{k} \rightarrow 0_{k} \quad \text{ with rate } B_{k} ,	\\
		0_{k} \rightarrow 1_{k} \quad \text{ with rate } D_{k} .  \\
		\end{cases}$
	\end{tabular}
\end{center}
In the bulk, $1_{k}$'s hop forward (resp. backward) with rate 1 (resp. $q$). Hence the relevant boundary parameters are $a_{k}:= \kappa^{+}_{ A_{k}, G_{k}}$, and $b_{k}:= \kappa^{+}_{ B_{k}, D_{k}}$. Since we need to consider all colourings at the same time, a typical point in the phase diagram for the mpASEP is described by the coordinates $\left( a_{1}, \ldots, a_{r}, b_{1}, \ldots, b_{r} \right) $.

In spite of a large number of parameters, finding the phases simplifies considerably because $a_{i} < a_{j}$ and $b_{j} < b_{i}$ always hold true for $i<j$. In order to describe the phases, each phase is identified with a word consisting of two letters $\vvmathbb{i}, \vvmathbb{j} \in \left\lbrace \vvmathbb{0}, \ldots, \vvmathbb{r} \right\rbrace $ with $0\leq i+j \leq r$. Further, it is convenient to use the shorthand notations $\overline{\vvmathbb{j}} =\vvmathbb{r} - \vvmathbb{j}$, and $\vvmathbb{j}_{\pm} = \vvmathbb{j} \pm \vvmathbb{1}$. Then the phases are as enlisted below. 

\begin{itemize}
	\item \underline{Phase $\phasel{0} \left(  0 \leq l < r  \right) $}: $\max \left\lbrace a_{r-l}, b_{1} \right\rbrace  < 1 < a_{r-l+1} $.	
	
	\item \underline{Phase $\phase{0}{r}$}: $\max\left\lbrace  1, b_{1} \right\rbrace <a_{1} $.
	
	\item \underline{Phase $\phasel{j}   \left( 0 < j < r, 0\leq l < r - j \right) $}: $\max\left\lbrace  a_{r-l}, b_{j+1} \right\rbrace < 1 < \min \left\lbrace a_{r-l+1} , b_{j}  \right\rbrace  $.
	
	\item \underline{Phase $\phaseol{j}{j}   \left( 0 < j < r \right) $}: $\max\left\lbrace 1, a_{j-1}, b_{j} \right\rbrace < \min \left\lbrace  a_{j}, b_{j-1} \right\rbrace $.
	
	\item \underline{Phase $\phase{r}{0}$}: $\max\left\lbrace  a_{r}, 1 \right\rbrace < b_{r} $.
\end{itemize}
The notation is such that one can easily keep track of the effect of $k$-colouring on each phase. To be more specific, we consider phase $\phase{i}{j}$. When we project the mpASEP onto the ASEP with the application of $k$-colouring, this phase maps to the HD phase for $1 \leq k\leq i$, to the LD phase for $r-j < k \leq r $, and to the MC phase for $i<k \leq r-j$.

We visualize all phases on a two-dimensional plane that passes through each phase. We focus on one such plane that is determined by 
\begin{equation}
	\begin{aligned}
		& a_{r} = \mu_{r-1} a_{r-1} = \cdots = \mu_{1} a_{1} \ ,  \\
		& b_{1} = \nu_{1} b_{2} = \cdots = \nu_{r-1} b_{r} \ ,
	\end{aligned}
\end{equation}
where we fix scalars $\mu_{i}, \nu_{i} > 1$ with $\mu_{i} > \mu_{j}$ and $\nu_{j} > \nu_{i}$ for $i,j \in [r-1]$ and $i<j$. On this plane, let $L_{i} \left(\text{resp. } R_{i}\right) $ be the point where the hyperplanes $a_{i} = 1 \ (\text{resp. } b_{i} = 1 )$ and $b_{i} = 0 \ (\text{resp. } a_{i} = 0)$ intersect. Thus the coordinates of $L_{i}$ and $R_{i}$ are given by
\begin{equation}
	\begin{aligned}
		& L_{i} \equiv 
				\begin{cases}
					\left( \mu_{i}/\mu_{1}, \ldots , \mu_{i}/\mu_{i-1}, 1, \mu_{i}/\mu_{i+1}, \ldots , \mu_{i}/\mu_{r-1}, \mu_{i}, 0, \ldots, 0 \right) \ \ \text{ if } 1 \leq i<r 
					\\
					\left( 1/\mu_{1}, \ldots ,  1/\mu_{r-1}, 1, 0, \ldots, 0 \right) \ \ \text{ if }  i = r , 
				\end{cases} \\
		& R_{i} \equiv 
				\begin{cases}
					\left( 0, \ldots, 0 , 1, 1/\nu_{1}, \ldots ,  1/\nu_{r-1} \right) \ \ \text{ if }  i = 1 , 
					\\
					\left( 0, \ldots, 0 , \nu_{i-1}, \nu_{i-1}/\nu_{1}, \ldots , \nu_{i-1}/\nu_{i-2}, 1, \nu_{i-1}/\nu_{i}, \ldots , \nu_{i-1}/\nu_{r-1} \right) \ \ \text{ if } 1 < i \leq r 
				\end{cases} \\
	\end{aligned} \label{eq:lrpoints}
\end{equation}
In the $\left( a_{1},\ldots, a_{r}\right) $-subspace, we define $a:= \left( \sum_{i=1}^{r} a_{i}^{2}\right) ^{1/2}$ which is the radial distance from the origin; likewise, we have $b:= \left( \sum_{i=1}^{r} b_{i}^{2}\right) ^{1/2}$ for $\left( b_{1},\ldots, b_{r}\right) $-subspace. We describe the structure of the phase diagram in Figure~\ref{fig:gpd} in terms of the parameters $a$ and $b$ as well as $L_{i}$'s and $R_{i}$'s. The phases $ \vvmathbb{j}\overline{\vvmathbb{j}}$ and $ \vvmathbb{j}_{-}\overline{\vvmathbb{j}_{-}}$ coexist on the semi-infinite line $1<a_{j}=b_{j}$ with $j \in \left[ r \right] $. All coexistence lines appear as thick lines in the phase diagram in Figure~\ref{fig:gpd}. In addition, we show the semi-infinite lines $1<a_{i}=b_{j}$ for $i>j$ as dashed lines. Unlike the coxistence lines, these lines do not separate different phases. The significance of these lines will be highlighted in the next subsection where we discuss densities and currents.

\begin{figure}[htbp!]
	\begin{center}
		\begin{tikzpicture}[scale=0.33]
		\draw[-{Latex[length=3.5mm]},thick] (0,0)--(0,36);
		\draw[-{Latex[length=3.5mm]},thick] (0,0)--(36,0);
		\draw[black,thick] (0,0) rectangle (32,4);
		\draw[black,thick] (0,0) rectangle (28,8);
		\draw[black,thick] (0,0) rectangle (24,11);
		\draw[black,thick] (0,0) rectangle (20,15);
		\draw[black,thick] (0,0) rectangle (16,20);
		\draw[black,thick] (0,0) rectangle (10.5,24);
		\draw[black,thick] (0,0) rectangle (7,28);
		\draw[black,thick] (0,0) rectangle (3.5,32);
		
		\draw[black,line width = 1.1pt] (32, 4)--(36,4.5);
		\draw[black,line width = 1.1pt] (28, 8)--(36,10.28571);
		\draw[black,line width = 1.1pt] (24, 11)--(36,16.5);
		\draw[black,line width = 1.1pt] (20, 15)--(36,27);
		\draw[black,line width = 1.1pt] (16, 20)--(28.8, 36);
		\draw[black,line width = 1.1pt] (10.5, 24)--(15.75,36);
		\draw[black,line width = 1.1pt] (7, 28)--(9, 36);
		\draw[black,line width = 1.1pt] (3.5, 32)--(3.9375, 36);
		
		\draw[dashed,black,thin] (24, 8)--(36, 12);
		\draw[dashed,black,thin] (20, 11)--(36, 19.8);
		\draw[dashed,black,thin] (16, 15)--(36, 33.75);
		\draw[dashed,black,thin] (7, 24)--(10.5, 36);
		\draw[dashed,black,thin] (10.5, 15)--(25.2, 36);
		
		\node at (38.2, 4.5) {\small$a_{1} = b_{1}$};
		\node at (39.14, 10.28571) {\small$a_{j-1} = b_{j-1}$};
		\node at (38.6, 12) {\small$a_{j} = b_{j-1}$};
		\node at (38.2, 16.5) {\small$a_{j } = b_{j}$};
		\node at (38.6, 19.8) {\small$a_{j +1} = b_{j}$};
		\node at (39.14, 27) {\small$a_{j +1} = b_{j+1}$};
		\node at (39.14, 33.75) {\small$a_{j+2}= b_{j+1}$};
		\node at (31.8, 36.8) {\small$a_{j+2}=b_{j+2}$};
		\node at (25.2, 36.8) {\small$a_{r-l}=b_{j+1}$};
		\node at (18, 36.8) {\small${a_{r-l}=b_{r-l}}$};
		\node[align = center] at (11.8, 33.8) {\small$a_{r-l+1}$\\ \ \ \ \ \small$=b_{r-l}$};
		\node at (9, 36.8) {\small${a_{r-l+1}=b_{r-l+1}}$};
		\node at (3, 36.8) {\small$a_{r}=b_{r}$};

		\node at (0,-0.75) {0};\node at (-0.75,0) {0};
		\node at (36.75,0) {\small$a$};\node at (0,36.8) {\small$b$};
		\node at (32.2,-0.8) {\small$L_{1}$};
		\node at (30.2,-0.8) {$\ldots$};
		\node at (28.2,-0.8) {\small$L_{j-1}$};
		\node at (24.2,-0.8) {\small$L_{j}$};
		\node at (20.4,-0.8) {\small$L_{j+1}$};
		\node at (16.4,-0.8) {\small$L_{j+2}$};
		\node at (13.5,-0.8) {$\ldots$};
		\node at (11,-0.8) {\small$L_{r-l}$};
		\node at (7.8,-0.8) {\small$L_{r-l+1}$};
		\node at (5.1,-0.8) {$\ldots$};
		\node at (3.6,-0.8) {\small$L_{r}$};
		
		\node at (-0.9, 4) {\small$R_{1}$};
		\node at (-0.9, 6) {$ \vdots $};
		\node at (-1.4, 8) {\small$R_{j-1}$};
		\node at (-0.9, 11) {\small$R_{j }$};
		\node at (-1.4, 15) {\small$R_{j+1}$};
		\node at (-1.4, 20) {\small$R_{j+2}$};
		\node at (-0.9, 22) {$ \vdots $};
		\node at (-1.4, 24) {\small$R_{r-l}$};
		\node at (-1.8, 28) {\small$R_{r-l+1}$};
		\node at (-0.9, 30) {$ \vdots $};
		\node at (-0.9, 32) {\small$R_{r}$};
		
		\node at (1.75, 2) {$ \phase{0}{0}$};
		\node at (5.25, 2 ) {$ \ldots$};
		\node at (8.75, 2) {$ \phasel{0}$};
		\node at (13.25, 2 ) {$ \ldots$};
		\node at (18, 2) {$ \vvmathbb{0}\overline{\vvmathbb{j}_{+}} $};
		\node at (22, 2) {$ \phaseol{0}{j} $};
		\node at (26, 2) {$  \vvmathbb{0}\overline{\vvmathbb{j}_{-}} $};
		\node at (30, 2 ) {$ \ldots$};
		\node at (34, 2 ) {$ \phase{0}{r}$};

		\node at (1.75, 6) {$ \vdots$};
		\node at (5.25, 6) {$ \ddots$};
		\node at (8.75, 6) {$ \vdots$};
		\node at (13.5, 6) {$ \ddots$};
		\node at (18, 6) {$ \vdots$};
		\node at (22, 6) {$ \vdots$};
		\node at (26, 6) {$\vdots$};
		\node at (33, 7) {$\ddots$};
		
		\node at (1.75,9.5) {$ \vvmathbb{j}_{-}\vvmathbb{0}$};
		\node at (5.25,9.5) {$ \ldots$};
		\node at (8.6, 9.5) {$ \vvmathbb{j}_{-}\ell$};
		\node at (13.25, 9.5) {$ \ldots$};
		\node at (18, 9.5) {$ \vvmathbb{j}_{-} \overline{\vvmathbb{j}_{+}}$};
		\node at (22, 9.5) {$ \vvmathbb{j}_{-} \overline{\vvmathbb{j}}$};
		\node at (31, 12.6) {$ \vvmathbb{j}_{-} \overline{\vvmathbb{j}_{-}}$};

		\node at (1.75,13) {$ \phase{j}{0}$};
		\node at (5.25,13) {$ \ldots$};
		\node at (8.75, 13) {$ \vvmathbb{j}\ell$};
		\node at (13.25, 13) {$ \ldots$};
		\node at (18, 13) {$ \vvmathbb{j} \overline{\vvmathbb{j}_{+}}$};
		\node at (28, 18) {$ \phaseol{j}{j}$};

		\node at (1.75, 17.5) {$ \vvmathbb{j}_{+} \vvmathbb{0}$};
		\node at (5.25, 17.5) {$ \ldots$};
		\node at (8.75, 17.5) {$ \vvmathbb{j}_{+}\ell$};
		\node at (22, 23) {$ \vvmathbb{j}_{+} \overline{\vvmathbb{j}_{+}}$};
		\node at (13.25, 17.5) {$ \ldots$};
		
		\node at (1.75, 26) {$ \overline{\ell}\vvmathbb{0}$};
		\node at (5.25, 26) {$ \ldots$};
		\node at (10.4, 30) {$ \overline{\ell}\ell$};
		
		\node at (1.75, 22) {$ \vdots$};
		\node at (5.25, 22) {$ \ddots$};
		\node at (8.75, 22) {$ \vdots$};
		\node at (15, 25) {$ \ddots$};
		
		\node at (1.75, 30) {$ \vdots $};
		\node at (6, 32) {$ \ddots $};
		\node at (1.75, 34) {$ \phase{r}{0}$};
		\end{tikzpicture}
		
		\caption{The schematic plot of a two-dimensional plane that shows all phases of the mpASEP. The plane is fixed by the scalars $\left( \mu_{1}, \ldots, \mu_{r-1}, \nu_{1}, \ldots, \nu_{r-1}  \right)$ in the $(a_{1},\ldots, a_{r},b_{1},\ldots, b_{r})$-space. We use the following shorthand notations: $\vvmathbb{j}_{\pm} \equiv \vvmathbb{j} \pm \vvmathbb{1}$, $\overline{\vvmathbb{j}} \equiv \vvmathbb{r}-\vvmathbb{j}$, and $\overline{\vvmathbb{j}_{\pm}} \equiv \vvmathbb{r}-\vvmathbb{j}_{\pm}$.}
		\label{fig:gpd}
	\end{center}
\end{figure}

\subsection{Currents and densities}
All densities and currents can be computed using the $k$-colouring. We discuss the calculations in details for phase $\phase{0}{j}$, and mention only the results for rest of the phase regions below. For our convenience, we define $f(x) = 1/(1+x)$, $\overline{f}(x) = 1- f(x),$ and $g(x,y) = (1-q) \left|  f(x) \overline{f}(x) - f(y)\overline{f}(y) \right| $. We summarize the results in Table~\ref{table:cdmpasep}.

\subsubsection*{\textbf{Phases $\phasel{0}$ and $\phase{0}{r}$}} The $k$-colourings project phase $\phasel{0}$ onto the MC phase for $k \leq r-l $ and LD phase for $k> r-l$. Therefore densities satisfy 
\begin{equation}
\begin{aligned}
& \sum_{i = 0}^{k-1} \rho_{i} = \frac{1}{ 2 } = \sum_{i = k}^{r} \rho_{i} , \quad 1 \leq k \leq r-l  , \\
& \sum_{i = 0}^{k-1} \rho_{i} = \overline{f}(a_{k}) = 1 - \sum_{i = k}^{r} \rho_{i}, \quad r-l < k \leq r. \\
\end{aligned}
\end{equation} 
From these relations, we have $\rho_{0}= 1/2$, and $ \rho_{i} = 0 $ for $0<i<r-l$. Thus each species $i$ greater than $0$ and smaller than $(r-l)$ are dynamically expelled from the system. Other species have positive densities as listed in Table~\ref{table:cdmpasep}.  Similarly, for the currents we have 
\begin{equation}
\begin{aligned}
& - \sum_{i = 0}^{k-1} J_{i} = g(1,0) = \sum_{ i = k }^{ r } J_{i}, \quad 1 \leq k \leq r-l  , \\
& - \sum_{i = 0}^{k-1} J_{i} = g( a_{k},0) = \sum_{ i = k }^{ r } J_{i}, \quad r-l < k \leq r  .
\end{aligned}
\end{equation} 
Straightforward calculation shows $ J_{0} = - g(1,0) $, $ J_{i} = 0 $ for $0<i<r-l$, whereas all species $i \geq r-l$ have positive currents: $J_{r-l} = g(1, a_{r-l+1})$, $J_{i} = g(a_{i}, a_{i+1})$ for $r-l < i < r $, and $J_{r} = g(a_{r}, 0)$. 

For the phase $\phase{0}{r}$, all colourings map to LD phase. Each species have nonzero densities as well as currents. Except species $r$, every species has negative currents (see Table~\ref{table:cdmpasep}).

See Figure \ref{fig:fig1} (a), (b) and (c) for density profiles obtained from numerical simulations in phases $\phase{0}{0}$, $\phase{0}{1}$ and $\phase{0}{2}$ for the case $r=2$.

\begin{table}[htbp!]
	\renewcommand*{\arraystretch}{1.3}
	\begin{center}
		\begin{tabular}{|c|c|c|c|}
			\hline 
			Phase &  Species  & Density $\rho$ & Current $J$  \\ \hline
			\multirow{5}{*}{$\phasel{0}$} 
			& $0$ & $ f(1)$ & $-g(1, 0)$\\ \cline{2-4}
			& $0< i < r - l$ & $ 0$ & 0\\ \cline{2-4}
			& $r-l$& $f(1) - f(a_{r-l+1}) $ & $g(1, a_{r-l+1})$ \\ \cline{2-4}
			& $r-l< i <r$ & $ f(a_{i}) - f(a_{i+1}) $ & $g(a_{i}, a_{i+1})$\\ \cline{2-4} 
			& $r$ & $ f(a_{r})$ & $g(a_{r},0)$ \\
			\hline
			\multirow{3}{*}{$\phase{0}{r}$} 
			& $0$ & $  \overline{f}( a_{1})  $ &  $ - g( a_{1},0)$\\ \cline{2-4}
			& $ 0<i<r $ & $ f(a_{i}) - f(a_{i+1}) $ &  $ g( a_{i},  a_{i+1}) $ \\ \cline{2-4}
			& $r$& $ f( a_{r})$ & $g( a_{r}, 0) $ \\
			\hline
			\multirow{7}{*}{$\phasel{j}$}
			& $0$ & $ f(b_{1}) $ & $-g(b_{1},0)$ \\ \cline{2-4} 
			& $ 0 < i <  j$& $f(b_{i+1}) - f(b_{i})$ & $-g(b_{i+1},b_{i})$ \\ \cline{2-4}
			& $j$ & $ f(1) - f(b_{j})$ & $-g(1,b_{j})$\\ \cline{2-4}
			& $ j < i < r-l$& $0$ & $0$ \\ \cline{2-4}
			& $r-l $& $ f(1) - f( a_{r-l+1}) $ & $g(1, a_{r-l+1}) $ \\ \cline{2-4}
			& $r-l < i < r$& $ f(a_{i}) - f(a_{i+1})$ & $ g(a_{i}, a_{i+1}) $ \\\cline{2-4}
			& $r$& $ f(a_{r})$ & $ g(a_{r},0) $ \\
			\hline
			\multirow{5}{*}{$\vvmathbb{j}\overline{\vvmathbb{j}}$} 
			& $ 0 $ & $ f(b_{1}) $ & $-g(b_{1}, 0)$ \\ \cline{2-4}
			& $ 0 < i < j $ & $ f( b_{i+1} ) - f( b_{i} ) $ & $ -g(b_{i+1}, b_{i})$ \\ \cline{2-4}
			& $ j $ & $ \overline{f}( a_{j+1} ) - f(b_{j})$ & $ \textrm{sign}( a_{j+1} - b_{j} ) g(a_{j+1}, b_{j}) $ \\ \cline{2-4}
			& $ j < i <r$ & $ f( a_{i} ) - f( a_{i+1} ) $ & $g(a_{i}, a_{i+1})$ \\ \cline{2-4}
			& $r$ & $ f(a_{r})$ & $g( a_{ r }, 0) $ \\ \cline{2-4}
			\hline
			\multirow{3}{*}{$\phase{r}{0}$} 
			& $0$& $ f(b_{1})$ & $-g(b_{1},0)$ \\ \cline{2-4}
			& $ 0 < i <  r$ & $ f(b_{i+1}) - f(b_{i})$ & $- g(b_{i+1}, b_{i})$ \\ \cline{2-4}
			& $r$& $ \overline{f}(b_{r})$ & $g(b_{r}, 0)$ \\ 
			\hline
		\end{tabular}
	\end{center}
	\vspace{0.33cm}	
	\caption{Bulk densities and currents in each phase for the mpASEP. We use the convention that $J>0$ (resp. $J<0$) if flux of particles is directed towards the right (resp. left) boundary.} 
	\label{table:cdmpasep}
\end{table}

\subsubsection*{\textbf{Phases $\phasel{j}$ and $\phaseol{j}{j}$}} Let us first consider the phase $\phasel{j}$ (with $\vvmathbb{j}>\vvmathbb{0}$ and $\ell < \overline{\vvmathbb{j}}$). All species $i$ with $j<i<r-l$ are dynamically expelled in this phase. These species have vanishing densities and currents. All other species have low densities. Any species equal to or smaller (resp. greater) than $j$ (resp. $r-l$) has negative (resp. positive) current. 

The phase $\phaseol{j}{j}$ differs from other phases of the form $\phasel{j}$ because no $k$-colouring projects phase $\phaseol{j}{j}$ onto the MC phase. No species is dynamically expelled from the system. Every species other than species $j$ has low density everywhere in this phase. The species $j$ assumes high density where $a_{j+1}$ and $b_{j}$ satisfies $f(a_{j+1}) + f(b_{j}) < 1/2$. Otherwise, density of the $j$'s is low as well. It must be noted, nevertheless, that $\rho_{j}$ varies continuously with the parameters $a_{j+1}$ and $b_{j}$ throughout phase $\phaseol{j}{j}$. The currents are nonzero for all species with the following exception. There is no current of species $j$ on the part of the hyperplane $a_{j+1} = b_{j}$ contained in phase $\phaseol{j}{j}$. This part of the phase diagram is the semi-infinite line $1<a_{j+1} = b_{j}$ shown as dashed line in Figure~\ref{fig:gpd}. To explain vanishing current of the $j$'s on this line, we note that $\sum_{i=0}^{j-1} \rho_{i} = f(b_{j})$ and $\sum_{i=j+1}^{r} \rho_{i} = f(a_{j+1})$ in phase $\phaseol{j}{j}$. Since correlations are expected to be absent in the theormodynamic limit, we have 
\begin{equation}
	J_{j} = (1-q) \rho_{j} \left( \sum_{i=0}^{j-1} \rho_{i} \right)  - (1-q) \left(  \sum_{l=j+1}^{r} \rho_{l} \right)  \rho_{j}.  \label{eq:currentj}
\end{equation}  
Thus $J_{j} $ becomes zero for $a_{j+1} = b_{j}$. Furthermore, it immediately follows that $J_{j}$ is positive (resp. negative) for $b_{j} < a_{j+1}$ (resp. $b_{j} > a_{j+1}$) in phase $\phaseol{j}{j}$. 

The simulation results for density profiles in phases $\phase{1}{0}$ and $\phase{1}{1}$ related to the case $r=2$ are recorded in Figure \ref{fig:fig1} (d), (e), (f) and (g).

\subsubsection*{\textbf{Phase $\phase{r}{0}$}} 
Every colouring projects phase $\phase{r}{0}$ onto the HD phase. All species have nonzero densities, as well as currents. All but species 0 have positive currents. 

The density profiles in phase $\phase{2}{0}$ for $r=2$ are plotted in Figure \ref{fig:fig1} (h).

\subsubsection*{\textbf{$\vvmathbb{j}_{-}\overline{\vvmathbb{j}_{-}}-\phaseol{j}{j}$ Coexistence Line and Other Semi-infinite Lines}}
The $k$-colouring maps $\vvmathbb{j}_{-}\overline{\vvmathbb{j}_{-}}-\phaseol{j}{j}$ coexistence line to the HD phase for $k<j-1$, to the LD phase for $k>j-1$, and to the coexistence line of the ASEP for $k=j-1$. All species have nonzero bulk densities, but the density profiles of species $(j-1)$ and $j$ are phase-segregated. These two species have linear density profiles (see Figure~\ref{fig:fig1}~(i) and (j) for simulation results related to the case $r=2$).

It is interesting to note that, on the semi-infinite line $1<a_{i}=b_{j}$ with $i>j$, the densities and currents always satisfy
\begin{equation}
	\begin{aligned}
		& \sum_{m=j+1}^{i} \rho_{m} = 1 - 2 f(a_{i}) \, , \qquad  \sum_{m=0}^{j} \rho_{m} = \sum_{n=i+1}^{r} \rho_{n} = f(a_{i}) \, , \\
		& \sum_{m=j+1}^{i} J_{m} = 0. 
	\end{aligned} \label{eq:sldensity}
\end{equation}
We have already discussed the densities and currents for $i=j+1$. For $i>j+1$, the line $1<a_{i}=b_{j}$ passes through more than one phase. Thus, the functional forms of the densities and currents for a point on this line will depend on the phase to which that point belongs.

\section{The generalized shock picture in the mpASEP}\label{sec:shock}
We describe here the generalized shock picture that we use to explain density profiles in each phase for mpASEP. It is convenient to consider first the shock picture on the coexistence lines in the phase diagram. Below, we will start with the coexistence lines, and then go on to discuss other phases.

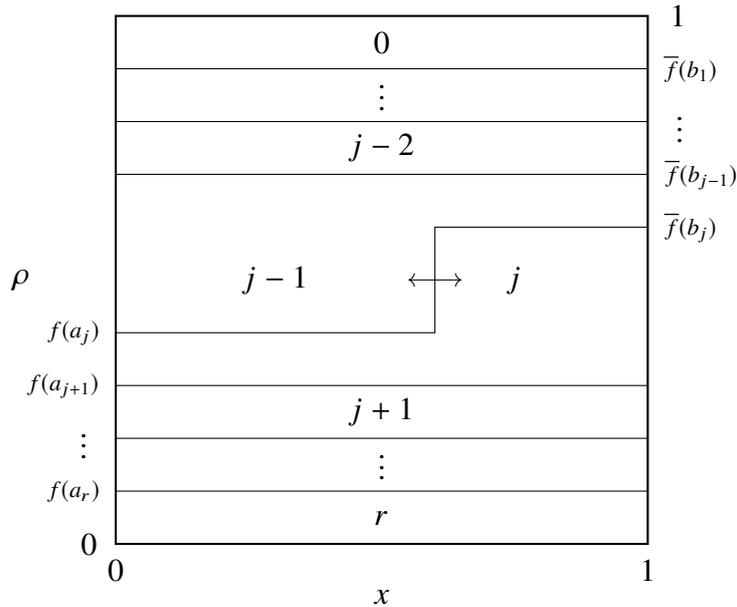
\begin{figure}[htbp!]
	\begin{center} 
		\begin{tikzpicture}[scale=0.7]
		\draw[black,thick] (0,0) rectangle (10,10);
		\draw[black] (0,4)--(6,4)--(6,6)--(10,6);
		\draw[black] (0,3)--(10,3);
		\draw[black] (0,1)--(10,1);
		\draw[black] (0,2)--(10,2);
		\draw[black] (0,7)--(10,7);
		\draw[black] (0,8)--(10,8);
		\draw[black] (0,9)--(10,9);
		
		\node[] at (-0.8,4){$\scriptstyle f(a_{j})$};
		\node[] at (-1,3){$\scriptstyle f(a_{j+1})$};
		\node[] at (-0.8,1){$\scriptstyle f(a_{r})$};
		\node[] at (11,7){$\scriptstyle\overline{f}(b_{j-1})$};
		\node[] at (10.8,6){$\scriptstyle\overline{f}(b_{j})$};
		\node[] at (10.8,9){$\scriptstyle\overline{f}(b_{1})$};
		\node[] at (-0.6,2){$\vdots$};
		\node[] at (10.6,8){$\vdots$};
		
		\node[] at (5,7.5){$j-2$}; 
		\node[] at (5,9.5){$0$}; 
		\node[] at (5,0.5){$r$};
		\node[] at (5,2.5){$j+1$};
		\node[] at (5,8.6){$\vdots$};
		\node[] at (5,1.6){$\vdots$};
		\node[] at (3,5){$j-1$};
		\node[] at (7.5,5){$j$};
		\draw[<->] (5.5,5)--(6.5,5);
		
		\node at (-1.8,5) {$\rho$};
		\node[] at (0,-0.5){$0$};\node[] at (10,-0.5){$1$};
		\node[] at (-0.5,0){$0$}; \node[] at (10.57,10){1};
		\node[] at (5,-1){$x$};
		
		\end{tikzpicture}
	\end{center}
	\caption{The generalized shock picture for the mpASEP on the $\vvmathbb{j}_{-}\overline{\vvmathbb{j}_{-}}-\phaseol{j}{j}$ coexistence line in the phase diagram in Figure~\ref{fig:gpd}. The schematic plot shows a shock formed between species $j-1$ and $j$. For $r=1$, one recovers the shock picture for the ASEP shown in Figure~\ref{fig:asepshock} from the generalized shock picture.  }
	\label{fig:gshockp}
\end{figure}

\subsubsection*{\textbf{$\vvmathbb{j}_{-}\overline{\vvmathbb{j}_{-}}-\phaseol{j}{j}$ Coexistence Line}} A shock between species $(j-1)$ and $j$ forms on the boundary of the phases $\vvmathbb{j}_{-} \overline{\vvmathbb{j}_{-}}$ and $\phaseol{j}{j}$ as shown in Figure \ref{fig:gshockp}.  Particles of every other species play the role of spectators, and maintain constant density in the system. The shock undergoes a random walk with zero net drift. Thus, the density profiles for $(j-1)$ and $j$ are linear, whereas $\rho_{i}$ is constant for all other species. For $r=1$, there is no other species except the 0's and 1's which take part in the shock. In this case, it is easy to verify that the generalized shock picture reduces to the one displayed in Figure~\ref{fig:asepshock}. 

\begin{figure}[htbp!]
	\centering
	\includegraphics[width = 0.8\textwidth]{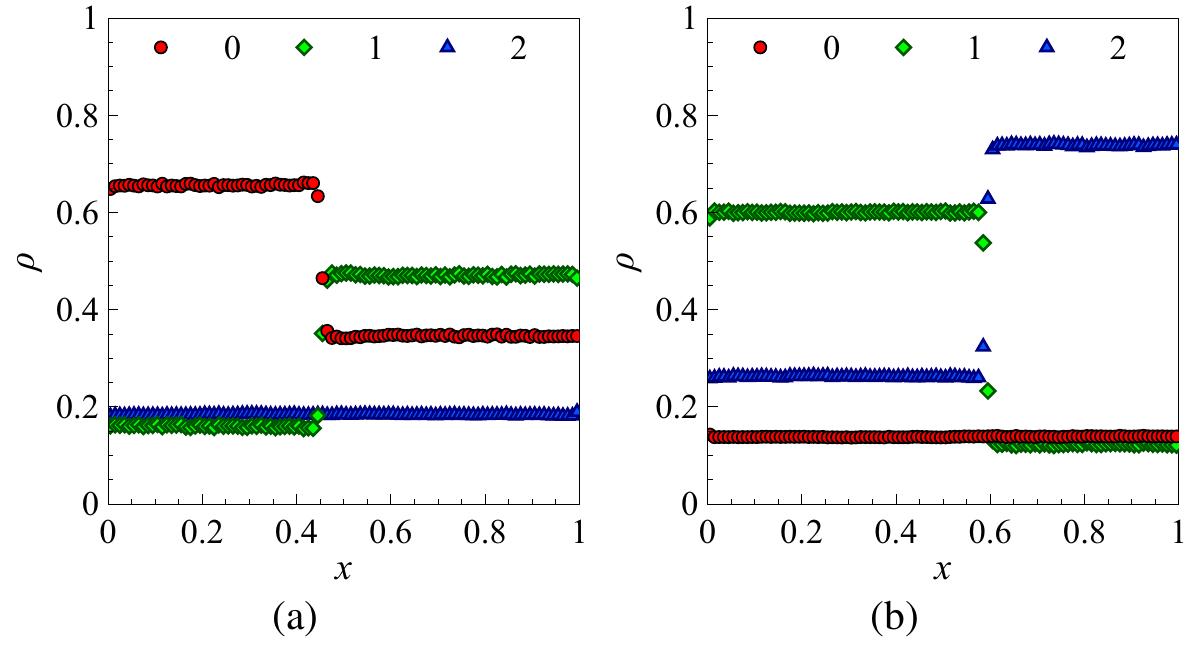}
	\caption{Densities on the phase coexistence regions in the mpASEP for $r=2$: the instantaneous density profiles for species 2 (blue triangles),
		1 (green diamonds), and 0 (red circles) on (a) $\phase{0}{1}-\phase{1}{1}$ coexistence region ($a_{1} \simeq 1.95 ,a_{2} \simeq 4.48 ,b_{1} \simeq 1.95 ,b_{2} \simeq 1.16 $), and (b) $\phase{1}{1}-\phase{2}{0}$ coexistence region ($a_{1} \simeq 1.1, a_{2} \simeq 2.84,b_{1} \simeq 6.21,b_{2} \simeq 2.84 $), where we fixed the lattice size to be 10000. \label{fig:shocks}}
\end{figure}

In Figure \ref{fig:shocks} (a) and (b), we plot instantaneous density profiles for $r=2$ obtained from numerical simulations pertaining to the coexistence regions $\phase{0}{2}-\phase{1}{1}$ and $\phase{1}{1}-\phase{2}{0}$ in the phase diagram. These simulations agree with the shock picture explained here.

\subsubsection*{\textbf{Phase $\vvmathbb{j}_{-} \overline{\vvmathbb{j}_{-}}$ and $\phaseol{j}{j}$}} As one moves away from $\vvmathbb{j}_{-}\overline{\vvmathbb{j}_{-}}-\phaseol{j}{j}$ coexistence line and approaches phase $\phaseol{j}{j}$, the $(j-1)-j$ shock acquires negative drift. Consequently, the shock gets pinned to the left boundary leading to the bulk densities for phase $\phaseol{j}{j}$ listed in Table~\ref{table:cdmpasep}. The total density of all species $i \geq j$ is always higher than $1/2$ in phase $\phaseol{j}{j}$. However, in phase $\vvmathbb{j}_{-} \overline{\vvmathbb{j}_{-}}$, the $(j-1)-j$ shock has positive velocity which pins the shock to the right. 
Thus, densities are such that the inequality $ \sum_{n=j}^{r} \rho_{n} < 1/2$ holds true in phase $\vvmathbb{j}_{-} \overline{\vvmathbb{j}_{-}}$.


\subsubsection*{\textbf{Phase $\phasel{j}$}} We consider all phases $\phasel{j}$ with $0 \leq j <r $ and $ 0< l <r-j $. First, we consider the phase $\vvmathbb{j} \overline{\vvmathbb{j}_{+}}$. This phase can be accessed as we move along the $\vvmathbb{j} \overline{\vvmathbb{j}} - \vvmathbb{j}_{+} \overline{\vvmathbb{j}_{+}}$ coexistence line towards the origin. On this coexistence line, the shock is formed between species $j$ and $j+1$. As one reaches phase $\vvmathbb{j} \overline{\vvmathbb{j}_{+}}$, the height of the shock reduces to zero along with $\sum_{m=0}^{j} \rho_{m} = 1/2 = \sum_{n=j+1}^{r} \rho_{n}$. All densities remain unchanged in this process except for species $j$ and $j+1$. This is because $k$-colouring maps phase $\vvmathbb{j}\overline{\vvmathbb{j}_{+}}$ to the MC phase rather than the coexistence line of the ASEP for $k=j+1$, whereas the projections remain same for $k \neq j+1$.

\begin{figure}[htbp!]
	\begin{center}
		\begin{tikzpicture}[scale=0.32]
		\draw[-{Latex[length=3.5mm]},thick] (0,0)--(0,24);
		\draw[-{Latex[length=3.5mm]},thick] (0,0)--(36,0);
		\draw[black,thick] (0,0) rectangle (32,4);
		\draw[black,thick] (0,0) rectangle (24,11);
		\draw[black,thick] (0,0) rectangle (20,15);
		\draw[black,thick] (0,0) rectangle (16,15);
		\draw[black,thick] (0,0) rectangle (10.5,15);
		\draw[black,thick] (0,0) rectangle (7,15);
		\draw[black,thick] (0,0) rectangle (3.5,20);
		
		\draw[black,line width = 1.1pt] (32, 4)--(36, 4.5);
		\draw[black,line width = 1.1pt] (24, 11)--(36,16.5);
		\draw[black,line width = 1.1pt] (20, 15)--(32,24);
		\draw[black,line width = 1.1pt] (3.5, 20)--(4.2,24);
		
		\draw[densely dotted, red, line width = 1.1pt] (0, 8.8)--(36, 8.8); 
		
		\draw[dashed,black,thin] (24, 4)--(36, 6);
		\draw[dashed,black,thin] (10.5, 11)--(22.91, 24);
		\draw[dashed,black,thin] (20, 11)--(36, 19.8);

		\node at (38.2, 4.5) {\small$a_{j} = b_{j}$};
		\node at (38.6, 6) {\small$a_{j+1} = b_{j}$};
		\node at (39.14, 16.5) {\small$a_{j+1 } = b_{j+1}$};
		\node at (39.14, 19.8) {\small$a_{j +2} = b_{j+1}$};
		\node at (32.5, 24.8) {\small$a_{j +2} = b_{j+2}$};
		\node at (5.5, 24.8) {\small$a_{r} = b_{r}$};
		\node at (38, 8.8) {\small$ b_{j} = C$};
		\node at (22.91, 24.8){\small$a_{r-l} = b_{j+1}$};
		
		\node at (0,-0.75) {0};\node at (-0.75,0) {0};
		\node at (36.75,0) {\small$a$};\node at (0, 24.8) {\small$b$};
		\node at (28,-0.8) {$\ldots$};
		\node at (32,-0.8) {\small$L_{j}$};
		\node at (24.4,-0.8) {\small$L_{j+1}$};
		\node at (20.4,-0.8) {\small$L_{j+2}$};
		\node at (16.4,-0.8) {\small$L_{j+3}$};
		\node at (13.5,-0.8) {$\ldots$};
		\node at (11,-0.8) {\small$L_{r-l}$};
		\node at (7.8,-0.8) {\small$L_{r-l+1}$};
		\node at (5.1,-0.8) {$\ldots$};
		\node at (3.6,-0.8) {\small$L_{r}$};
		
		\node at (-0.9, 2) {$ \vdots $};
		\node at (-0.9, 17.5) {$ \vdots $};
		\node at (-0.9, 4) {\small$R_{j}$};
		\node at (-1.4, 11) {\small$R_{j + 1}$};
		\node at (-1.4, 15) {\small$R_{j+2}$};
		\node at (-0.9, 20) {\small$R_{r}$};
		\node at (-0.9, 8.8){\small$R^{*}$};

		\node at (1.75, 2) {$ \vdots$};
		\node at (5.25, 2) {$ \ddots$};
		\node at (8.75, 2) {$ \vdots$};
		\node at (13.5, 2) {$ \ddots$};
		\node at (18, 2) {$ \vdots$};
		\node at (22, 2) {$ \vdots$};
		\node at (28, 2) {$\ddots$};
		\node at (34, 2.5) {$\ddots$};
		
		\node at (1.75,6.5) {$ \vvmathbb{j}\vvmathbb{0}$};
		\node at (5.25,6.5) {$ \ldots$};
		\node at (8.6, 6.5) {$ \vvmathbb{j}\ell$};
		\node at (13.25, 6.5) {$ \ldots$};
		\node at (18, 6.5) {$ \vvmathbb{j} \overline{\vvmathbb{j}_{++}}$};
		\node at (22, 6.5) {$ \vvmathbb{j} \overline{\vvmathbb{j}_{+}}$};
		\node at (31, 12.6) {$ \vvmathbb{j} \overline{\vvmathbb{j}}$};
		
		\node at (1.75,13) {$ \vvmathbb{j}_{+} \vvmathbb{0}$};
		\node at (5.25,13) {$ \ldots$};
		\node at (8.75, 13) {$ \vvmathbb{j}_{+} \ell$};
		\node at (14, 13) {$ \ldots$};
		\node at (18, 13) {$ \vvmathbb{j}_{+} \overline{\vvmathbb{j}_{++}}$};
		\node at (28, 18) {$ \vvmathbb{j}_{+} \overline{\vvmathbb{j}_{+} }$};

		\node at (1.75, 22) {$ \phase{r}{0}$};
		\node at (1.75, 17.5) {$ \vdots $};
		\node at (11, 20) {$ \ddots $};
		\end{tikzpicture}
		
		\caption{The scematic plot of the hyperplane $b_{j} = C$ (red dotted line) that passes through all phases of the form $\phasel{j}$. Here, we use the notation $\vvmathbb{j}_{++} = \vvmathbb{j} + \vvmathbb{2}$. Some details of the phase diagram are omitted in this plot. We have included only the essential parts that make this illustration suitable to our discussion regarding phase $\phasel{j}$ in Section~\ref{sec:shock}.}
		\label{fig:hplane}
	\end{center}
\end{figure}
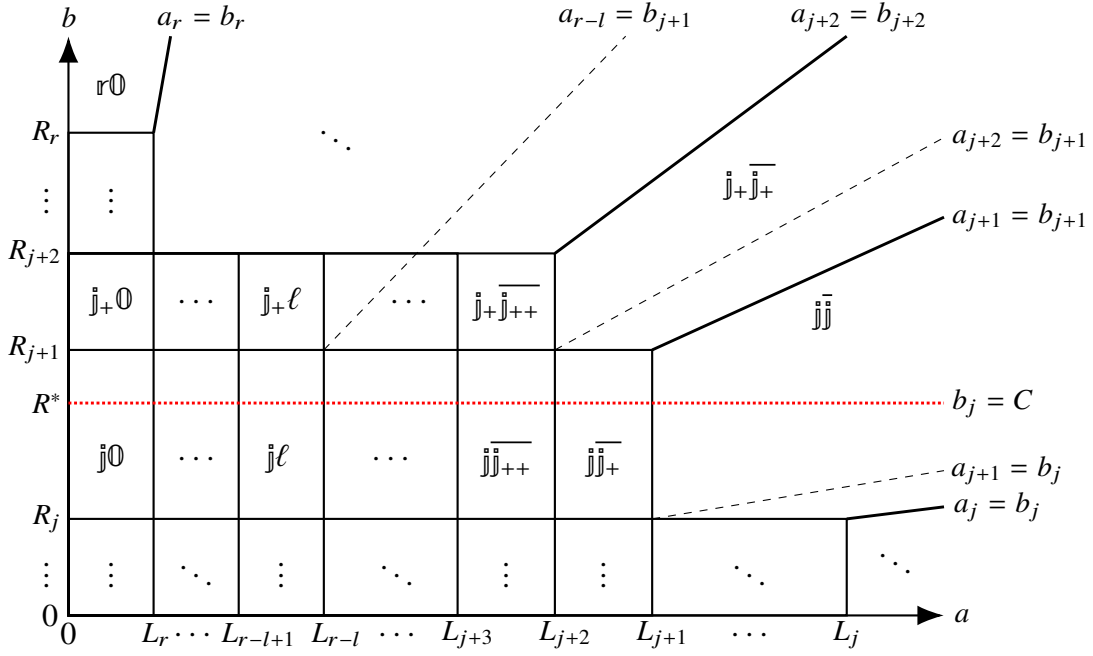

Before we discuss density profiles in the rest of the phases, it is useful to understand the dynamical expulsion of every species greater than $j$ and lower than $(r-l+1)$ in phase $\phasel{j}$ with $\ell < \overline{\vvmathbb{j}_{+}}$. Let us consider the simplest case $\ell = \overline{\vvmathbb{j}_{++}}$ where $\vvmathbb{j}_{++} =\vvmathbb{j} + \vvmathbb{2}$. We consider the semi-infinite line $ 1<a_{j+2} = b_{j+1} $  on which we have $\rho_{j+1}= 1-2f(b_{j+1})$. As one moves along this part of the phase diagram towards the origin, $\rho_{j+1}$ decreases as $b_{j+1}$ decreases. Eventually, density of the $(j+1)$'s becomes zero for $b_{j+1} \leq 1$ when one enters the phase $\vvmathbb{j} \overline{\vvmathbb{j}_{++}}$ from phase $\vvmathbb{j}_{+} \overline{\vvmathbb{j}_{+}}$. Moreover, we observe that $ \sum_{m=0}^{j} \rho_{m} = \sum_{n=j+2}^{r}\rho_{n}=1/2$ is satisfied in phase $\vvmathbb{j} \overline{\vvmathbb{j}_{++}}$. Although $\rho_{j+1}$ vanishes in phase $\vvmathbb{j} \overline{\vvmathbb{j}_{++}}$, it is possible that the $(j+1)$'s enter the lattice because of nonzero boundary rates. To see why such particles are driven out of the system, we must consider boundary effects near the left and right boundaries. The densities satisfy $\sum_{m=0}^{j} \rho_{m}(x) < 1/2 < \sum_{n=j+2}^{r}\rho_{n}(x)$ and $\sum_{m=0}^{j} \rho_{m}(x) >1/2 > \sum_{n=j+2}^{r}\rho_{n}(x)$ at the normalized site position $x$ near the left and right boundary respectively due to boundary effects. Thus the $(j+1)$'s will be driven away from the bulk at both the boundaries. Similarly, we can understand dynamical expulsion in phase $\phasel{j}$ by considering the semi-infinite line $1<a_{r-l} = b_{j+1}$. Indeed, one checks using \eqref{eq:sldensity} that $\sum_{n=j+1}^{r-l} \rho_{n}$ vanishes as one approaches phase $\phasel{j}$ along this line.

It suffices to understand density profiles in all phases $\phasel{j}$ for fixed $\vvmathbb{j}$. First, we fix a constant $C \in (1, \nu_{j}/\nu_{j-1})$, and the hyperplane $b_{j} = C$ on which $b_{j+1}<1<b_{j}$ is satisfied. One such hyperplane that passes through the phases under consideration here is indicated as the red dotted line in Figure~\ref{fig:hplane}. The coordinate of $R^{*}$ where the hyperplanes $b_{j} =C$ and $a_r =0 $ intersect on the two-dimensional phase diagram is given by $ R^{*} \equiv C R_{j}$.

Now, we explain the density profiles in phase $\vvmathbb{j} \overline{\vvmathbb{j}_{++}} $ starting with its adjacent phase $\vvmathbb{j}\overline{\vvmathbb{j}_{+}}$ for which we have already explained the density profiles using the shock picture. As one moves along the hyperplane $b_{j} = C$ from phase $\vvmathbb{j}\overline{\vvmathbb{j}_{+}}$ towards phase $\vvmathbb{j}\overline{\vvmathbb{j}_{++}}$, $\rho_{j+2}$ increases to $(f(1)-\overline{f}(a_{j+3}))$ ($f(1)$ for $j+2=r$) at the hyperplane $a_{j+2} =1$ which is the boundary between these phases. The same expression for density of species $(j+2)$ remains unaltered throughout phase $\vvmathbb{j}\overline{\vvmathbb{j}_{++}}$. In constrast, the density of the $(j+1)$'s decreases to zero. The $(j+1)$'s stay dynamically expelled in phase $\vvmathbb{j}\overline{\vvmathbb{j}_{++}}$ as explained previously. However, other species retain their expressions for densites in this process. 

In general, as we move further along $ b_{j} = C $ towards phase $\phase{j}{0}$, $\rho_{n}$ vanishes and $\rho_{n+1}$ becomes $(f(1)-f(a_{n+2}))$ ($f(1)$ for $n=r-1$) at the hyperplane $a_{n+1}=1$ for $j+1<n<r$. To reach $\phasel{j}$ starting from $\vvmathbb{j}\overline{\vvmathbb{j}_{+}}$ along the line $ b_{j} = C $, one must pass through the hyperplanes $a_{n} =1$ for $j+1 < n \leq r-l$. In this process, $\rho_{i}$ declines to zero for $j<i<r-l$, and $\rho_{r-l}$ becomes $(f(1)- f(a_{r-l+1})$ ($f(1)$ for $l=0$), where functional forms of other densities remain the same as in phase $\vvmathbb{j}\overline{\vvmathbb{j}_{+}}$.

\section{Conclusion}
We have studied the phase diagram for the mpASEP, a multispecies asymmetric simple exclusion model with arbitrary number of species. A salient feature of the model is that it is endowed with boundaries permeable to all species in the system. The dynamics at the boundaries is so defined that we can project the multispecies model onto the single-species ASEP model by dint of colouring. As mentioned earlier, we do not have a matrix ansatz for this model to evaluate the physical quantities directly. It will be a challenging and immensely interesting problem to find such an exact solution for the mpASEP.  

\subsection*{Acknowledgement} I thank my PhD supervisor Prof. Arvind Ayyer for discussions regarding this project, critical reading of the manuscript, as well as many helpful suggestions. This work is supported by Indian Institute of Science, Bengaluru.

\bibliographystyle{plain}
\bibliography{mpASEP}

\end{document}